\documentclass[acmsmall,screen]{acmart}
\AtBeginDocument{%
  }

\setcopyright{acmlicensed}
\copyrightyear{2024}
\acmYear{2024}
\acmDOI{XXXXXXX.XXXXXXX}

\acmConference[Conference acronym 'XX]{}{June 03--05,
  2018}{Woodstock, NY}
\acmISBN{978-1-4503-XXXX-X/18/06}

\newcommand{\code}[1]{\texttt{#1}}
\begin{document}

\title{A Survey of Recent Developments in SYCL Compiler Implementations}

\author{Huy Trinh}
\email{h3trinh@uwaterloo.ca}
\orcid{0003-4652-3870}
\affiliation{%
  \institution{University of Waterloo}
  \city{Waterloo}
  \country{Canda}}

\renewcommand{\shortauthors}{Huy et al.}

\begin{abstract}
This survey discusses recent advancements in SYCL compiler implementations, one of the crucial aspects of compiler construction for heterogeneous computing systems.  We explore the transition from traditional compiler construction, from Single-Source Multiple Compiler Passes (SMCP) to a more advanced approach to Single-Source Single Compiler Pass (SSCP). The survey analyzes multiple papers that researched the different developments of SYCL implementation based on SSCP and their approach to enhancing performance and addressing separate challenges.
\end{abstract}




\maketitle

\section{Introduction}
With the advent of data processing and Artificial Intelligence, many vendors have developed their hardware and domain-specific accelerators, thus, highlighting the need for portable code to be run on various hardware. As a result, a generic open standard solution software stack called SYCL \cite{syclWebsite}, is introduced to address these challenges. SYCL \cite{syclWebsite}, which was developed by Khronos Group and announced in March 2014, is a single source C++-based specification, that helps programmers dispatch their code work on massively parallel cores and leverage C++ features on a range of heterogeneous platforms, ranging from CPU, GPU, ASIC, FPGA, DSP and other accelerators.

On the one hand, there is an emerging trend in adopting SYCL code in many scientific and HPC applications \cite{10.1145/3388333.3388643} \cite{inproceedings}  \cite{Pascuzzi_2022}. On the other hand, SYCL also enables a portable way to run neural networks on various accelerator backends \cite{10.1007/978-3-031-40843-4_41}. The survey will first discuss SYCL's existing implementation of SYCL compilers and their portability evaluation, concentrating on three major implementations: OpenSYCL, DPC++ \cite{10.1145/3388333.3388653} and ComputeCpp \cite{ComputeCppGuide}. Then it goes through the modification of the compilation flow when SYCL is optimized on the CPU. After that, we will explain the difference between SMCP and SSCP compilation models, which pave the way for implementing the SYCL construct by combining a single-pass compiler and unified code presentation binary across various backends. Finally, we briefly introduce an MLIR-based SYCL compiler to address the limitations of C++ that SYCL has.

 
 

\section{SYCL Compiler Portability and Compilation Flow}
\begin{figure}[htbp]
  \centering
  \includegraphics[width=\linewidth]{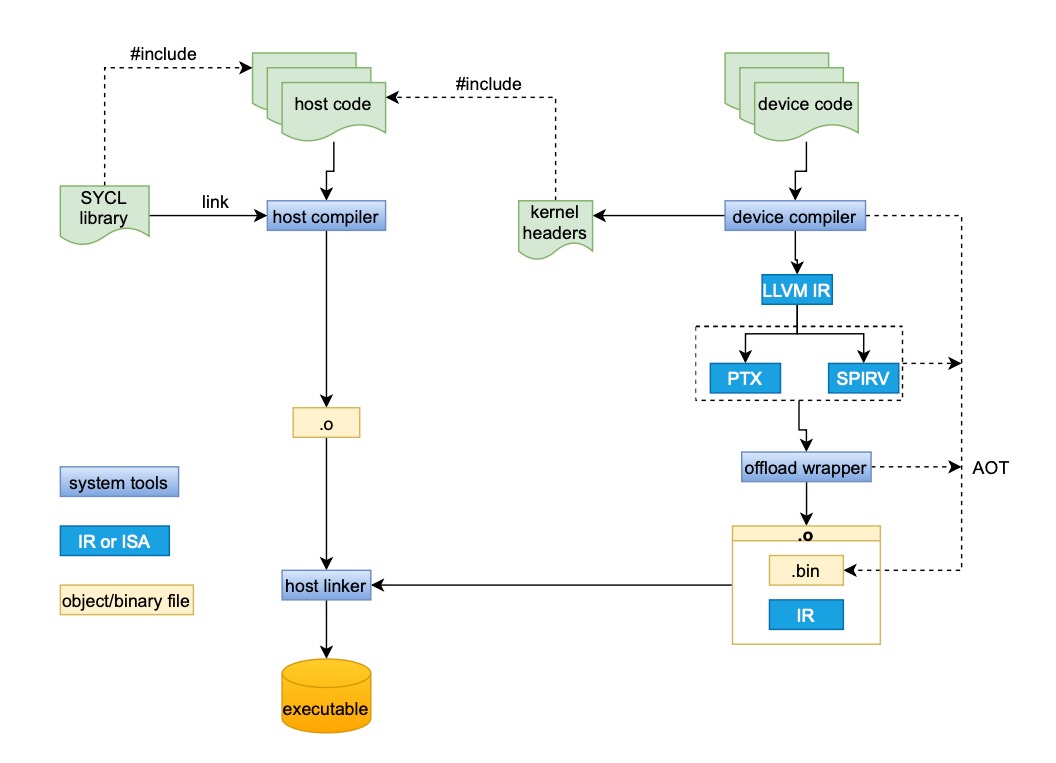}
  \caption{The overview of SYCL compilation. \cite{wang2021research}}
  \Description{The overview of SYCL compilation.}
  \label{fig:appfig1}
\end{figure}

Since SYCL is a single-source programming model (both CPU and accelerator source code are in the same C++ translation unit), the C++ Compiler can do type-checking and optimize host and device code together with interaction API such as kernel launches, and data transfer. The benefit of this single-source design allow the reuse same code on both hosts and devices. In contrast, OpenCL is a separate source model, which refers to different source codes between host and device. The early version of SYCL acts as a C++ abstraction layer over OpenCL, targeting making parallel programming easier based on OpenCL. Later, SYCL 2020 was transitioned as a generalized model and made OpenCL one of many different potential backends. SYCL comprises two-stage compilations: one for the host code and one for the device code. While the host code in the SYCL application can be compiled with any C++ compiler (host compiler), SYCL is designed to compile kernel code to suitable final targeted devices. This characteristic indicates Single-Source, Multiple Compiler Passes (SCMP) models (discussed more in \ref{sec:sscp_smcp}) where there is distinct management in compilation stages. Therefore, SYCL implementations must provide an SYCL-aware compiler, called the device compiler, such as DPC++ \cite{10.1145/3388333.3388653} and ComputeCPP \cite{ComputeCppGuide}, to compile kernels separately from host code, aligning with the method of generating intermediate presentation before final compilation. Thoman et al. \cite{Thoman2022Compilation}'s study evaluates the compiling speed of the SYCL program generated by their own code generator of these implementations. In addition, the study also examined other minor SYCL implementations, including hipSYCL and triYCL while tunning various import parameters such as the number of kernels, buffers and accessors used, the data types computed with, the number of loop nests, and the instruction mix inside the kernel. Figure \ref{fig:appfig1} schematically illustrates the SYCL compilation process, starting with host and device code that undergoes compilation through host and device compilers. SYCL library components and kernel header are fed into the device compiler, producing intermediate representations like LLVM IR, PTX and SPIR-V, which are then processed through an offload wrapper before finally generating the executable. Intel DPC++ compiler, in this case, links together and wraps the generated device image in the host object that will be read later by SYCL runtime.

Study \cite{PENNYCOOK20131439} shows deep analysis of SYCL Compiler implementations capability of code portability and their performance evaluation. The research emphasizes how SYCL has evolved and constructed to address challenges of performance portability across various hardware platforms. Three implementations of SYCL compiler include Open SYCL, DPC++ \cite{10.1145/3388333.3388653} and CoputeCpp \cite{ComputeCppGuide} were examined and their compilers optimize the translation from C++ into executable without locking developers into vendor-specific ecosystem. A vital aspect discussed is their ability to integrate more compilation strategies different from the normal traditional model, which requires separate passes for host and device code. This alternative approach, Single-Source Single Compiler Pass (SSCP), which will be discussed later in section \ref{sec:sscp_smcp}, shows the effectiveness of optimization considering the interaction between host and device code from the outset of the compilation process in the study. However, for some complex applications ( for example, those involving finite element methods or finite difference calculation), SYCL compilers struggle to match performance compared to native approaches like CUDA or MPI \cite{Thoman2022Compilation}. This indicates while SYCL provides excellent portability, there is sometimes a trade-off for performance. Nguyen et al. \cite{Nguyen_2023}, otherwise, present SYCL's potential to outperform CUDA implementation in porting batched iterative solver (which is essential in complex simulations like plasma physics) onto Intel GPU compared to NVIDIA hardware. The transition of these solvers to SYCL leverages its portability to other different architectures, including CPUs, GPUs, and FPGA. Although the role of SYCL's single-source approach in cross-platform reuse is crucial, we observed that performance may be sacrificed for portability in scenarios where we would like to utilize maximum hardware usage by specifically fine tunning according to characteristics of applications.

\section{Improving Performance on CPU Architectures}
One development in SYCL compilation flow is optimized when the compilation target is on the CPU. There are various reasons to optimize CPU code or host code. Normally, when targeting the CPU, OpenCL introduces unnecessary significant overheads like memory I/O transmission from/to OpenCL device due to the inherent differences in handling device-specific and host code. Pietro et al. \cite{10.1145/3528425.3529099} propose a novel compilation flow of bypassing OpenCL backends and target CPU. 

In general, the SYCL compilation flow can target the CPU through two approaches: using OpenCL as the backend and through the SYCL host device. In a prior way, after extracting, the intermediate format is fed into the OpenCL device compiler to become a format that accelerators can execute. On platforms like x86, which can act as an accelerator, the device code is statically compiled offline, and the flexibility of supporting multiple devices is not required. Along with SPIR/SPIR-V, there are other intermediate supported formats such as PTX (NVIDIA platforms), GCN (AMD platforms), and even OpenCL C code. However, these intermediate formats introduce higher overhead than finalizing SPIR binaries as they contain OpenCL-parsing/analysis information. Another approach is to use any C++-compliant compiler to compile the SYCL program through the SYCL host device. The SYCL host device does not operate as an OpenCL device but as a native C++ implementation of a device, which manages various tasks such as queue management, thread dispatch, scheduling, and synchronization within the CPU itself.

\begin{figure}[htbp]
  \centering
  \includegraphics[width=\linewidth]{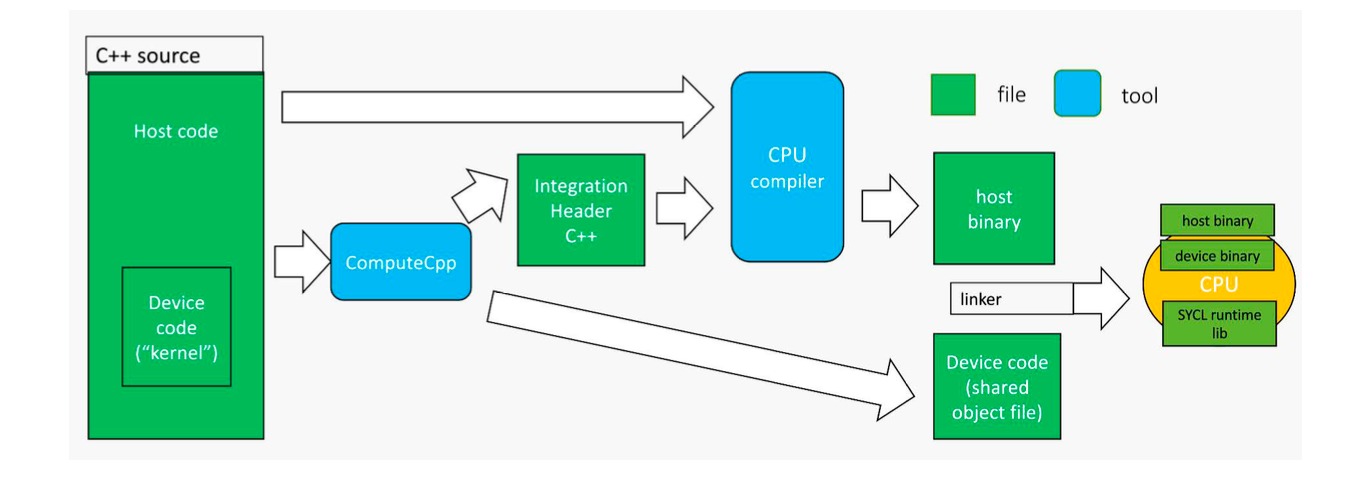}
  \caption{SYCL offline host compilation: The SYCL compiler directly compiles the device code into CPU code. \cite{10.1145/3528425.3529099}}
  \Description{A more detailed description of Figure 1.}
  \label{fig:fig2}
\end{figure}

While the former method needs additional translation stages that interpret and optimize device-specific IRs, the latter approach allows more direct integration with modern C++ Compiler and compilation into machine code of the host CPU, which reduces the latency typically associated with launching kernels on a computing device. Moreover, this integration will leverage more fine tunes targeting the architecture they support, including loop unrolling, function inlining, and advanced vectorization. Based on this, the authors \cite{10.1145/3528425.3529099} modified the normal SYCL host compilation flow by getting rid of the intermediate SPIR/SPIR-V/etc representation and using the compiler's backend information to compile kernel into the shared library as shown in figure \ref{fig:fig2}. However, the ComputeCpp SYCL compiler \cite{ComputeCppGuide} still undertakes the device code extraction step. The shared library is then linked to the application using a standard linker. Since the platform target is known and fixed, the device code can be statically compiled offline \cite{10.1145/3388333.3388652}, and multiple supported devices' flexibility is not needed. Finally, the executable is independent of any OpenCL implementations or other backends, in which kernels are executed in a Single Instruction Multiple Data (SIMD) fashion.

\begin{figure}[htbp]
  \centering
  \includegraphics[width=\linewidth]{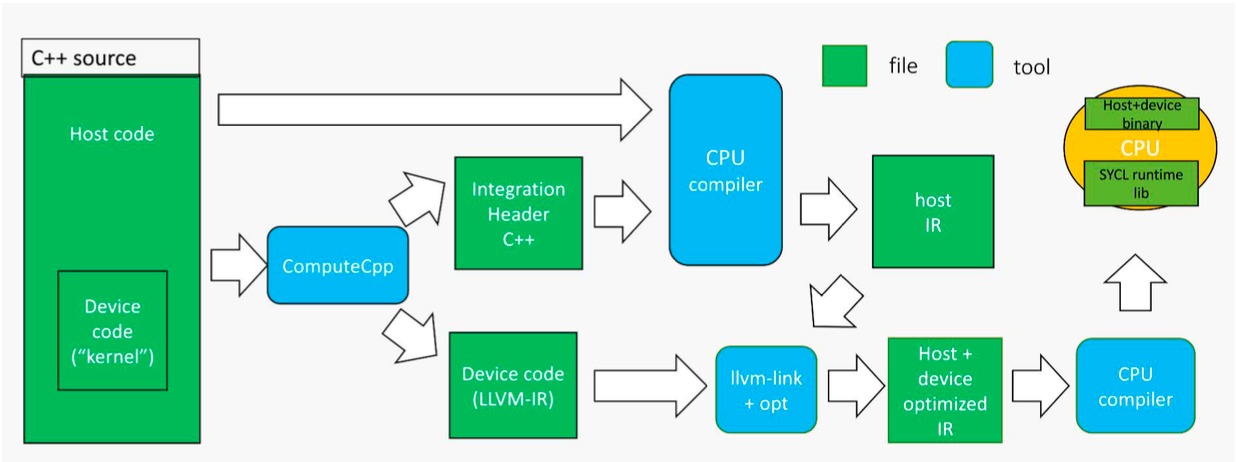}
  \caption{SYCL optimized offline host compilation. \cite{10.1145/3528425.3529099}}
  \Description{A more detailed description of Figure 1.}
  \label{fig:fig3}
\end{figure}

Figure \ref{fig:fig3} depicts authors' further optimization on hosts and device code separately but related entities under the same compiling environment. Compiler ComputeCPP \cite{ComputeCppGuide}, built on top of Clang, emits the kernel's intermediate presentation as LLVM IR to provide a platform for integrating more optimizations. For example, Whole Function Vectorization, which is usually applied externally, now can be implemented directly within the LLVM pass framework. In my observation, it is natural that there is a trade-off between general-purpose code and architecture-specific tunning, in this study case, CPU optimization. The study \cite{10.1145/3528425.3529099} provides methodologies proving this, however, fails to address the line between CPU optimization and maintaining the generality of SYCL applications.

\section{SSCP And SMCP Compilation Models}
\label{sec:sscp_smcp}
\begin{figure}[htbp]
  \centering
  \includegraphics[width=\linewidth]{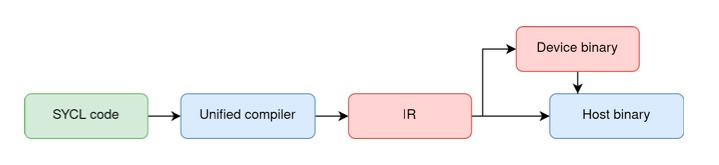}
  \caption{The single-source, single compiler pass (SSCP) model with late outlining. \cite{10.1145/3585341.3585351}}
  \Description{A more detailed description of Figure 1.}
  \label{fig:fig4}
\end{figure}

\begin{figure}[htbp]
  \centering
  \includegraphics[width=\linewidth]{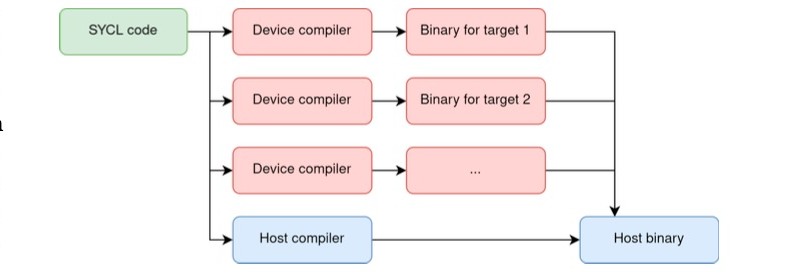}
  \caption{The single-source, multiple compiler passes (SMCP) model. \cite{10.1145/3585341.3585351}}
  \Description{The single-source, multiple compiler passes (SMCP) model.}
  \label{fig:fig5}
\end{figure}

The figures~\ref{fig:fig4} and \ref{fig:fig5} highlight the significant difference between the Single-Source, Single Compiler Pass (SSCP) model and the traditional Single-Source, Multiple Compiler Passes (SMCP) model. The SSCP model consolidates both host and device compilation into a single pass, replacing preprocessor directives usage by integrating conditional logic into IR code. This is achieved through construct statements, such as \code{if target()} in nvc++, evaluated at lowering IR to machine code stage. Moreover, compared to the SMCP model, the SCCP model can potentially prevent errors introduced from complex and separate compilation paths. For instance, in SYCL, the issue relates to C++ lambdas which is widely used in kernel definition. This can lead to undefined behaviour when implementing different data layouts of lambda captures between host and device pass in SMCP implementation. However, the SSCP model ensures the consistency of data layout residing across host-device boundaries. Other SYCL constructs such as kernel lambda naming and device code linking are simplified to avoid subtle bugs arising from complex conditional compilations. In addition to maintaining the portability, enhancing the performance and reducing compile time, authors \cite{10.1145/3585341.3585351} believe SSCP's unified form establishes a foundation for future SYCL modern compiler construction aiming at next-generation computing architectures.  From our perspective, we believe these optimizations must be balanced with robust debugging tools to ensure the simplicity of
the compilation process while not over the cost of maintainability in the community.

\section{Single-Pass SYCL Compiler with Unified Code Representation}
\label{sec:single-pass}
Although intermediate presentations are designed for interchanged and portable format code, these are not yet universally supported by all hardware vendors. For example, SPIR-V ingestion is not currently supported by the NVIDIA OpenCL and AMD. In addition, there is a need for universal binaries that can be distributed among users when their hardware configurations are unknown. Extended from hip SYCL, the authors \cite{10.1145/3585341.3585351} present the new compiler following the Single-source, single compiler pass (SSCP) model combined with the unified code presentation across backends. This new compiler comprises modifications in two phases.

In the initial compile-time phase, the compiler extracts and stores the LLVM IR of the kernel into the host binary's hardware-independent format. This encapsulation enables the universal capability of compiled code over various hardware configurations without backend-specific adaption. In detail, the process involves sophisticated techniques such as outlining kernel entry points and reachable, further refined through IR constants. These are global, non-constant C++ variables that will be transformed into constant LLVM IR at a later stage. These constants include flags like \code{is\_device} (to determine whether code is running on a device) and code {hcf\_content} to hold embedded device image. In general, LLVM IR which is used for device offload, would use additional address space annotations that describe memory type referenced by pointers. However, this generic generated presentation does not use specific address spaces for pointers and maintains the universal code representation. Moreover, this generic presentation does not bind to any specific backend in terms of the SYCL builtin at this stage. These will be replaced later during the next stage.

During the runtime stage, this embedded IR is lowered into supported backend's driver format such as PTX for NVIDIA, SPIR-V for Vulkan-compliant and amdgcn for AMD devices in \cite{10.1145/3585341.3585351}. To facilitate this, the author integrates llvm-to-backend infrastructure into hipSYCL, enhancing the system's extension and enabling new targets to the compilation pipeline to efficiently adapt to a range of backend specifications. Initially, the embedded HCF (hipSYCL Container Format) containing prior device IR is loaded. Based on the list of exported and imported symbols provided in HCF data, undefined \code{SYCL\_EXTERNAL} functions have to be resolved to maintain the integrity of the code across different modules and libraries. Following this, now the IR constants providing target backend and device information are set and guide subsequent optimizations and transformations. After these steps, optimized LLVM IR is finally transformed into a format depending on specific backends. Currently, the paper only supports CUDA backend, Intel GPU and AMD GPU. Their approach, however, increases the compilation time by approximately 20\% longer compared to the regular compilation. This increment is a trade-off for achieving flexibility and hardware independence at runtime. The authors argue that this compelling additional time is offset by eliminating the need for multiple target-specific compilations. Normally, this process may consume more time and resources when compiling over distinct hardware platforms.

\section{MLIR-based SYCL Compiler}
However, since SYCL is based on C++, it does have some limitations as a C++-based language when comes to compiler optimization. An example
would be the loss of high-level program structure and domain-specific in-
formation in the early stage of the compilation pipeline. On the other hand, MLIR (Multi-Level Intermediate Representation) \cite{lattner2020mlir} compiler framework can address these issues through its dialect by allowing high-level presentation of structures. Although paper \cite{10.1145/3585341.3585351} states the IR constants provide the optimization and transformation guild for device code in section \ref{sec:single-pass}, these take place completely in isolation from host compilation. Therefore, it prevents passing relevant information into the device optimization pipeline such as the context of the device kernel. To overcome this, \cite{tiotto2023experiences} uses MLIR  compiler framework \cite{lattner2020mlir} to capture SYCL  semantics on the high level of abstraction, bridging the gap between high-level abstract syntax tree representation of C++ to more regular LLVM IR normally used in device compilation. Moreover, MLIR is capable of nesting operations, allowing both host and device code reasoning at the same time.  Similar to \cite{10.1145/3585341.3585351} which uses the Single-Source Single Compiler Pass (SSCP) approach, the authors \cite{tiotto2023experiences} take advantage of information sharing between host and device compiler in SSCP, featuring the host code transformation by analyzing device code and vice versa. Additionally, they use Polygeist \cite{9563011} as a device compiler to translate from C++ AST to MLIR. However, the author experienced a struggle to handle compulsory C++ constructs (for example virtual functions and exceptions) on the host side at first. They then alternatively get MLIR host code from LLVM IR to leverage device compilation. Thus, this method creates a joint presentation of host and device code within MLIR's nested IR structure, which sets a framework to optimize device code using insights obtained from host code analysis.

The development of the SYCL-MLIR compiler has crucially extended the MLIR framework capabilities to optimize compilers, including several static analyses and utilities. The alias analysis benefits from domain-specific knowledge embedded in SYCL operation, which plays a key role in optimizing memory operation and data flow within the SYCL application. This shows the possible integration of enhancement in alias analysis and reaching definition analysis within the SYCL dialect of MLIR. For instance, alias analysis in the SYCL-MLIR compiler collaborates with the SYCL dialect operations' semantics to ascertain easier non-aliasing cases. Also, reaching definition analysis helps track value modification accurately across the SYCL program flow. Based on accurate data dependency and usage patterns, this ensures informative communication and the compiler can decide on optimization.  The integration of MLIR within the SYCL compilation pipeline is a promising development that aligns with my understanding of modern compiler design.

\section{Conclusion}
In conclusion, the survey has provided a comprehensive examination of recent advancements in the SYCL compiler implementation, framing the discourse within the context of heterogeneous computing system's ongoing research. Our survey discusses the SSCP versus SMCP compilation model and how SSCP becomes the potential model for later development and simplifying workflow. However, this simplification should not detract from comprehensive compilation and debugging capability. The concept of a unified code presentation is laying the foundation for future development in portable binaries targeting more complex hardware architectures. In addition, the exploration into MLIR-based SYCL compiler showcases an innovative approach to addressing some limitations of current SYCL while preserving high-level program structure in the compilation pipeline.


\bibliographystyle{ACM-Reference-Format}
\bibliography{references}
\clearpage
\appendix 
\section*{Appendix}

\end{document}